\documentclass{elsart}
\usepackage{amsfonts}
\usepackage{amssymb}
\usepackage{amsmath}
\usepackage[dvips]{graphicx}

\begin{document}

\begin{frontmatter}

\title{Cryptanalyzing a discrete-time chaos synchronization secure communication system}

\author{G. \'{A}lvarez\corauthref{corr}},
\author{F. Montoya},
\author{M. Romera},
\author{G. Pastor}

\corauth[corr]{Corresponding author: Email: gonzalo@iec.csic.es}

\address{Instituto de F\'{\i}sica Aplicada, Consejo Superior de
Investigaciones Cient\'{\i}ficas, Serrano 144---28006 Madrid,
Spain}

\begin{abstract}
This paper describes the security weakness of a recently proposed
secure communication method based on discrete-time chaos
synchronization. We show that the security is compromised even
without precise knowledge of the chaotic system used. We also make
many suggestions to improve its security in future versions.

\end{abstract}

\end{frontmatter}

\section{Introduction}

In recent years, a growing number of cryptosystems based on chaos
have been proposed \cite{asocscs,cc}, many of them fundamentally
flawed by a lack of robustness and security
\cite{stusc,cocborcr,emmbc,uamccs,pwtcisea,bcsugse,bcscuas,ccscurm,coaces,coacscs,coaecc,otsoacespwccifcp,coadccuek}.
In~\cite{sdcudtcs}, a secure communication system based on chaotic
modulation using discrete-time chaos synchronization is proposed.
Two different schemes of message encoding are presented. In the
first scheme, the binary message ($m(i)=\pm 1$) is multiplied by
the chaotic output signal of the transmitter and then sent to
drive the receiver system. In the second scheme, the binary
message is modulated by multiplication with the chaotic output
signal and then is fed back to the transmitter system and
simultaneously sent to the receiver system.

Discrete-time chaotic systems are generally described by a set of
nonlinear difference equations. The first communication system
based on modulation by multiplication can be described by:

\begin{equation}\label{eq:modmultr}
\rm transmitter\left\{ \begin{array}{l}
 x_1(i + 1) = 1-\alpha x_1^2(i) + x_2(i) \\
 x_2(i + 1) = \beta x_1(i) \\
 s(i) = x_1(i) \cdot m(i) \\
 \end{array} \right.
\end{equation}

\begin{equation}\label{eq:modmulrc}
\rm receiver\left\{ \begin{array}{l}
 \hat{x}_1(i + 1) = 1-\alpha s^2(i) + \hat{x}_2(i) \\
 \hat{x}_2(i + 1) = \beta \hat{x}_1(i) \\
 \hat{m}(i)=s(i)/\hat{x}_1(i)
 \end{array} \right.
\end{equation}

The communication scheme using modulation by multiplication and
feedback, with a modification to avoid divergence due to feedback,
is described by:

\begin{equation}\label{eq:modmulfbtr}
\rm transmitter\left\{ \begin{array}{l}
 x_1(i + 1) = 1-\alpha (s(i)- \left\lfloor{\frac{s(i)+P}{2P}}\right\rfloor 2P)^2 + x_2(i) \\
 x_2(i + 1) = \beta x_1(i) + 0.05x_1(i)(m(i)-1) \\
 s(i) = x_1(i) \cdot m(i) \\
 \end{array} \right.
\end{equation}

\begin{equation}\label{eq:modmulfbrc}
\rm receiver\left\{ \begin{array}{l}
 \hat{x}_1(i + 1) = 1-\alpha (s(i)- \left\lfloor{\frac{s(i)+P}{2P}}\right\rfloor 2P)^2 + \hat{x}_2(i) \\
 \hat{x}_2(i + 1) =\beta \hat{x}_1(i)+0.05(s(i)-\hat{x}_1(i)) \\
 \hat{m}(i)=s(i)/\hat{x}_1(i)
 \end{array} \right.
\end{equation}

with $P=(1+\sqrt{6.6})/2.8$.

Although the authors seemed to base the security of their
cryptosystems on the chaotic behavior of the output of the Henon
non-linear dynamical system, no analysis of security was included.
It was not considered whether there should be a key in the
proposed system, what it should consist of, what the available key
space would be, what precision to use, and how the key would be
managed.

In the next section we discuss the weaknesses of this secure
communication system using the Henon attractor and make some
suggestions to improve its security.

\section{Attacks on the proposed system}
\label{sec:attack}

\subsection{The key space}

Although it is not explicitly stated in \cite{sdcudtcs}, it is
assumed that the key is formed by the two parameters of the map,
$\alpha$ and $\beta$. Thus, in \cite{sdcudtcs}, the key is fixed
to $k=\{\alpha,\beta\}=\{1.4,0.3\}$. However, in \cite{sdcudtcs}
there is no information given about what the key space is. The key
space is defined by all the possible valid keys. The size of the
key space $r$ is the number of encryption/decryption key pairs
that are available in the cipher system.

In this chaotic scheme the key space is nonlinear because all the
keys are not equally strong. We say that a key is \emph{weak} or
\emph{degenerated} if it is easier to break a ciphertext encrypted
with this key than breaking a ciphertext encrypted with another
key from the key space.

The study of the chaotic regions of the parameter space from which
valid keys, i.e., parameter values leading to chaotic behavior,
can be chosen is missing in \cite{sdcudtcs}. A possible way to
describe the key space might be in terms of positive Lyapunov
exponents. According to \cite[p. 196]{caitds}, let $\mathbf{f}$ be
a map of ${\mathbb{R}}^m$, $m\geq 1$, and
$\{{\mathbf{x}}_0,{\mathbf{x}}_1, {\mathbf{x}}_2,\dots\}$ be a
bounded orbit of $\mathbf{f}$. The orbit is chaotic if

\begin{enumerate}
    \item it is not asymptotically periodic,
    \item no Lyapunov exponent is exactly zero, and
    \item the largest Lyapunov exponent is positive.
\end{enumerate}

\begin{figure}[h]
\center \includegraphics{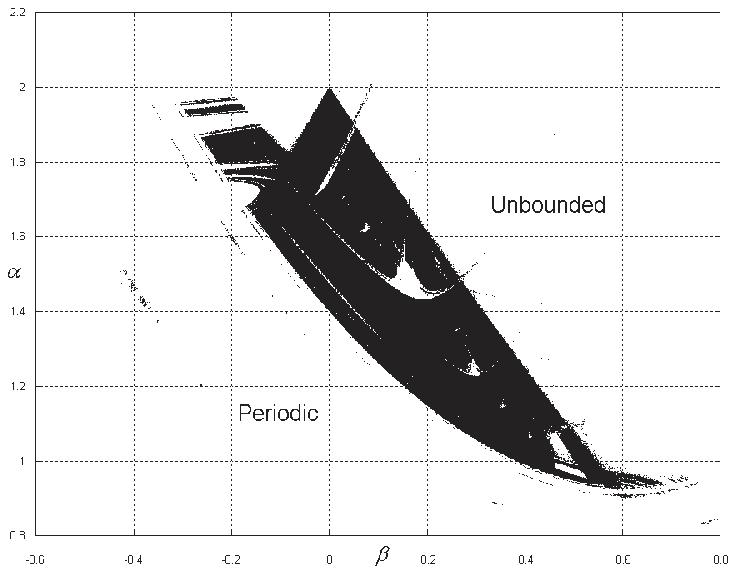} \caption{\label{fig:lyap}Chaotic
region for the Henon attractor.}
\end{figure}

The largest Lyapunov exponent can be computed for different
combinations of the parameters. If it is positive, then the
combination can be used as a valid key. In Fig.~\ref{fig:lyap},
the chaotic region for the Henon attractor used in \cite{sdcudtcs}
has been plotted. This region corresponds to the keyspace. In
general, parameters chosen from the lower white region give rise
to periodic orbits, undesirable because the ciphertext is easily
predictable. Parameters chosen from the upper white region give
rise to unbounded orbits diverging to infinity, and hence the
system can not work. Therefore, both regions should be avoided to
get suitable keys. Only keys within the black region are good. And
even within this region, there exist periodic windows, unsuitable
for robust keys.

This type of irregular and often fractal chaotic region shared by
most secure communication systems proposed in the literature is
inadequate for cryptographic purposes because there is no easy way
to define its boundary. And if the boundary is not mathematically
and easily defined, then it is hard to find suitable keys within
the key space. This difficulty in defining the key space
discourages the use of the Henon map. Instead, complete chaoticity
for any parameter value should be preferred. Piecewise linear
(PWL) maps are a good choice because they behave chaotically for
any parameter value in the useful interval
\cite{spodplcmatricaprc}.

\subsection{Insensitivity to parameter mismatch}

\begin{figure}[h]
\center \includegraphics{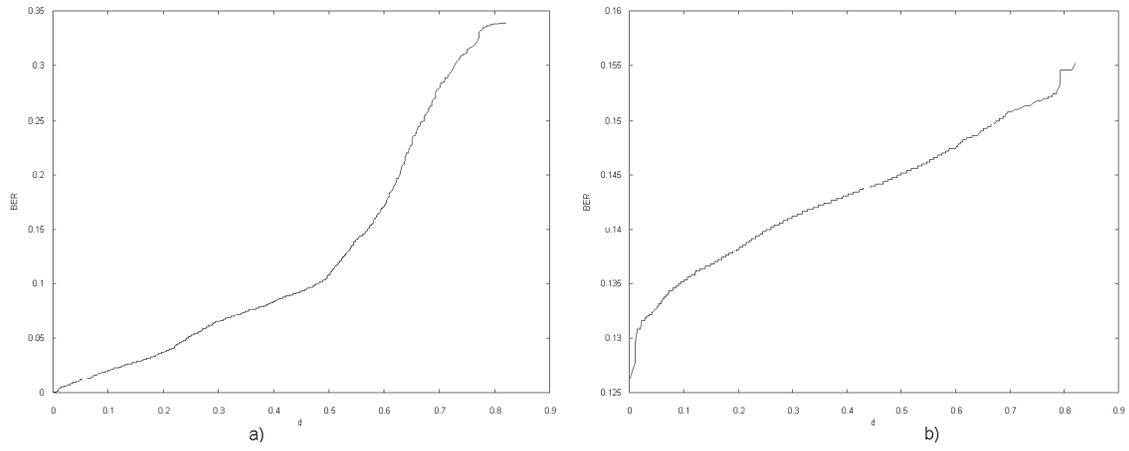} \caption{\label{fig:error}BER
when decrypting the ciphertext with a key at a distance $d$ from
the real encryption key, $k=\{\alpha,\beta\}=\{1.4,0.3\}$: (a)
modulation by multiplication; (b) modulation by multiplication and
feedback. Note the difference in scale.}
\end{figure}

Both communication systems, the one based on modulation by
multiplication and the one using modulation by multiplication and
feedback, can only have valid keys carefully chosen from the
chaotic region plotted in Fig.~\ref{fig:lyap} to avoid periodic
windows and divergence. Due to low sensitivity to parameter
mismatch, if the system key is fixed to
$k=\{\alpha,\beta\}=\{1.4,0.3\}$ as in \cite{sdcudtcs}, then any
key $k'$ chosen from the same key space will decrypt the
ciphertext into a message $m'$ with an error rate which is well
below 50\%. Fig.~\ref{fig:error} plots the bit error rate (BER)
when the ciphertext encrypted with
$k=\{\alpha,\beta\}=\{1.4,0.3\}$ is decrypted using keys $k'$ from
the valid key space at a distance $d$ from $k$. For this
experiment the Euclidean distance was chosen:

\begin{equation}\label{eq:d}
d=\sqrt{(\alpha-\alpha\,')^2+(\beta-\beta\,')^2}
\end{equation}

This insensitivity to parameter mismatch due to the coupling
between transmitter and receiver renders the system totally
insecure when the Henon map is used. A different map more
sensitive to small differences in the parameter values should be
used to grant security.

\subsection{Brute force attacks}

A brute force attack is the method of breaking a cipher by trying
every possible key. The quicker the brute force attack, the weaker
the cipher. Feasibility of brute force attacks depends on the key
space size $r$ of the cipher and on the amount of computational
power available to the attacker. Given today's computer speed, it
is generally agreed that a key space of size $r<2^{100}$ is
insecure.

However, this requirement might be very difficult to meet by this
cipher because the key space does not allow for such a big number
of different strong keys. For instance, Fig.~\ref{fig:lyap} was
created using a resolution of $10^{-3}$, i.e., there are
$1400\times 3000$ different points. To get a number of keys
$r>2^{100}\simeq 10^{30}$, the resolution should be $10^{-15}$.
However, with that resolution, thousands of keys would be
equivalent, unless there is a strong sensitivity to parameter
mismatch, which is usually lost by synchronization, even when
using a different chaotic map.

\subsection{Statistical analysis}

\begin{figure}[h]
\center \includegraphics{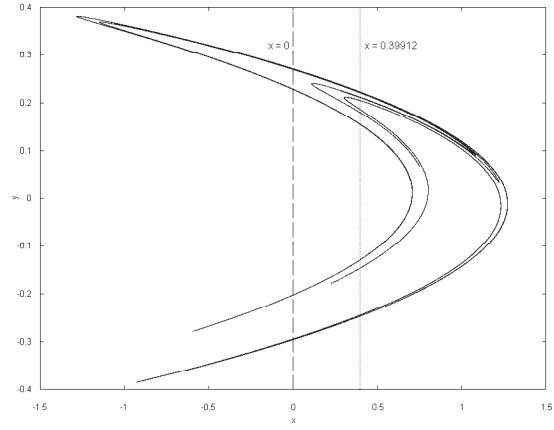} \caption{\label{fig:map}100,000
successive points obtained by iteration of the Henon map for
$\{\alpha,\beta\}=\{1.4,0.3\}$.}
\end{figure}

Fig.~\ref{fig:error}a shows that the error is upper bounded:
BER$\leq0.33$. This is a consequence of the fact that the orbit
followed by any initial point in the Henon attractor is not
uniformly distributed, because in average it spends two thirds of
the time above $x=0$. As a consequence, mixing the cleartext with
the output of a function whose probability density is not uniform
will result in a weak cryptosystem. In Fig.~\ref{fig:map} the
Henon attractor is plotted. It can be observed that the
distribution is far from flat because the orbit visits more often
the region $x>0$. In average, two thirds of the iterates lie to
the right of $x=0$ (depicted as a dashed line). This fact allows
the attacker to guess in average two thirds of the encrypted bits,
even with no knowledge about the transmitter/receiver structure.

To get a balanced distribution, the threshold should be moved to
the right \cite{thaaakg}. Let $x_m$ denote the real value such
that

\begin{equation}\label{eq:media}
    P(x_i\leq x_m)=P(x_i>x_m)=0.5.
\end{equation}

A good estimation presented in \cite{thaaakg} is
$\hat{x}_m=0.39912$, depicted as a dotted line in
Fig.~\ref{fig:map}. However, this result is difficult to apply
provided the way in which the Henon attractor is used by the
cryptosystem. Therefore, it is seen again that the Henon map is a
bad choice as a chaotic map for this communication scheme. A
different map with a balanced distribution, i.e., whose orbit
visits with equal frequency the regions above and below a certain
level $x=0$, should be chosen to prevent statistical attacks.

\subsection{Plaintext attacks}

In the previous sections we showed that the use of the Henon map
is not advisable because of its inability to define a good key
space, of its low sensitivity to parameter mismatch, and of its
non uniformly distributed orbits. We are to show next that if a
different map is used, the security of the communication system
will not improve if the same key is used repeatedly for successive
encryptions.

According to \cite[p. 25]{ctap}, it is possible to differentiate
between different levels of attacks on cryptosystems. In a known
plaintext attack, the opponent possesses a string of plaintext,
$p$, and the corresponding ciphertext, $c$. In a chosen plain
text, the opponent has obtained temporary access to the encryption
machinery, and hence he can choose a plain text string, $p$, and
construct the corresponding cipher text string, $c$.

The cipher under study behaves as a modified version of the
one-time pad \cite[p. 50]{ctap}. The one-time pad uses a randomly
generated key of the same length as the message. To encrypt a
message $m$, it is combined with the random key $k$ using the
exclusive-OR operation bitwise. Mathematically,

\begin{equation}\label{eq:pad}
c(i)=m(i)+k(i)\mod2,
\end{equation}

where $c$ represents the encrypted message or ciphertext. This
method of encryption is perfectly secure because the encrypted
message, formed by XORing the message and the random secret key,
is itself totally random. It is crucial to the security of the
one-time pad that the key be as long as the message and never
reused, thus preventing two different messages encrypted with the
same portion of the key being intercepted or generated by an
attacker.

Eq.~(\ref{eq:modmultr}) and Eq.~(\ref{eq:modmulfbtr}) are used to
generate a keystream
$\{x_1(1)=k(1),x_1(2)=k(2),x_1(3)=k(3),\ldots\}$. This keystream
is used to encrypt the plain text string according to the rule

\begin{equation}\label{eq:rule}
c(i)=k(i)\cdot m(i)
\end{equation}

Therefore, if the attacker possesses the plaintext $m(i)$ and its
corresponding ciphertext $c(i)$, he will be able to obtain $k(i)$.
If the same key, i.e. the same parameter values, is used to
encrypt any subsequent message in the future, it will generate an
identical chaotic orbit, which is already known. As a consequence,
when $c(i)$ and $k(i)$ are known in Eq.~(\ref{eq:rule}), $m(i)$ is
readily obtained by the attacker.

Obviously, when using this cryptosystem, regardless of the choice
of the chaotic map, the key can never be reused. A slight
improvement to partially enhance security even when the key is
reused consists of randomly setting the initial point of the
chaotic orbit at the transmitter end. Synchronization will
guarantee that the message is correctly decrypted by the
authorized receiver. However, an eavesdropper would have more
difficulty in using past chaotic orbits because they will diverge
due to sensitivity to initial conditions.

\section{Conclusions}
\label{sec:conclusion}

The proposed cryptosystem using the Henon map is rather weak,
since it can be broken without knowing its parameter values and
even without knowing the transmitter precise structure. However,
the overall security might be highly improved if a different
chaotic map with higher number of parameters is used. The
inclusion of feedback makes it possible to use many different
systems with non symmetric nonlinearity as far as the whole space
is folded into a bounded domain to avoid divergence. However, to
rigorously present future improvements, it would be desirable to
explicitly mention what the key is, how the key space is
characterized, what precision to use, how to generate valid keys,
and also to perform a basic security analysis. For the present
work \cite{sdcudtcs}, the total lack of security discourages the
use of this algorithm as is for secure applications.

\ack{This work is supported by Ministerio de Ciencia y
Tecnolog\'{\i}a of Spain, research grant TIC2001-0586. Our thanks
to Moez Feki for his comments and source code.}

\clearpage \pagestyle{empty}

\section*{Figures}

\clearpage

\clearpage

\end{document}